\documentclass[a4paper,11pt]{article}
\usepackage{jinstpub} % for details on the use of the package, please see the JINST-author-manual
\usepackage{lineno}

\usepackage{siunitx}

\proceeding{25$^{\text{th}}$ International Workshop on Radiation Imaging Detectors Workshop\\
30 June 2024 to 4 July 2024\\
Lisbon, Portugal}

\title{\boldmath Characterisation of analogue MAPS produced in
the 65 nm TPSCo process}

\author[a,b,1]{E. Ploerer,\note{Corresponding author.}}
\author[c]{H. Baba}
\author[d]{J. Baudot}
\author[d]{A. Besson}
\author[d]{S. Bugiel}
\author[e]{T. Chujo}
\author[d]{C. Colledani}
\author[d]{A. Dorokhov}
\author[d]{Z. El Bitar}
\author[d]{M. Goffe}
\author[c]{T. Gunji}
\author[d]{C. Hu-Guo}
\author[b]{A. Ilg}
\author[d]{K. Jaaskelainen}
\author[f]{T. Katsuno}
\author[g]{A. Kluge}
\author[h]{A. Kostina}
\author[d]{A. Kumar}
\author[b]{A. Lorenzetti}
\author[b]{A. Macchiolo}
\author[g]{M. Mager}
\author[e]{J. Park}
\author[e]{S. Sakai}
\author[d]{S. Senyukov}
\author[d]{H. Shamas}
\author[e]{D. Shibata}
\author[g]{W. Snoeys}
\author[h]{P. Stanek}
\author[g]{M. Suljic}
\author[h]{L. Tomasek}
\author[d]{I. Valin}
\author[f]{R. Wada}
\author[f]{Y. Yamaguchi}
\author[]{and the ALICE Collaboration}

\affiliation[a]{Vrije Universiteit Brussel, 1050 Brussels, Belgium}
\affiliation[b]{University of Zurich, 8057 Zurich, Switzerland}
\affiliation[c]{University of Tokyo, Tokyo 113-8654, Japan}
\affiliation[d]{Université de Strasbourg, CNRS, IPHC UMR 7178, F-67000 Strasbourg, France}
\affiliation[e]{University of Tsukuba, Tsukuba, Ibaraki 305-8577, Japan}
\affiliation[f]{Hiroshima University, Higashi-Hiroshima 739-8526, Japan}
\affiliation[g]{CERN, CH-1211 Geneva 23, Switzerland}
\affiliation[h]{Czech Technical University in Prague, 160 00 Prague, Czech Republic}

\emailAdd{eduardo.ploerer@cern.ch}

\abstract{Within the context of the ALICE ITS3 collaboration, a set of MAPS small-scale test structures were developed using the 65 nm TPSCo CMOS imaging process with the upgrade of the ALICE inner tracking system as its primary focus. One such sensor, the Circuit Exploratoire 65 nm (CE-65), and its evolution the CE-65v2, were developed to explore charge collection properties for varying configurations including collection layer process (standard, blanket, modified with gap), pixel pitch (15, 18, \SI{22.5}{\micro\meter}), and pixel geometry (square vs hexagonal/staggered).
In this work the characterisation of the CE-65v2 chip, based on $^{55}$Fe lab measurements and test beams at CERN SPS, is presented. Matrix gain uniformity up to the $\mathcal{O}$(5\%) level was demonstrated for all considered chip configurations. The CE-65v2 chip achieves a spatial resolution of under \SI{2}{\micro\meter} during beam tests. Process modifications allowing for faster charge collection and less charge sharing result in decreased spatial resolution, but a considerably wider range of operation, with both the \SI{15}{\micro\meter} and \SI{22.5}{\micro\meter} chips achieving over 99\% efficiency up to a $\sim$180 e$^{-}$ seed threshold. The results serve to validate the 65 nm TPSCo CMOS process, as well as to motivate design choices in future particle detection experiments. 
}


\begin{document}
\maketitle
\flushbottom

\section{Introduction}
\label{sec:intro}

Monolitihic Active Pixel Sensors (MAPS) combine the passive sensor and active readout chip onto the same silicon die. 
MAPS offer a variety of advantages with respect to their hybrid counterparts, including a lower material budget, cost, reduced power consumption, and smaller pitch as bump bonding is not needed. 
MAPS have already seen adoption by a wide variety of ongoing and proposed (collider) physics experiments including the STAR Experiment \cite{Dorokhov:2011zzb}, the ALICE Experiment \cite{ALICE:2013nwm}, the EIC \cite{Contin:2023hqf}, and the FCC-ee \cite{Barchetta:2021ibt}.

The ALICE experiment implemented a full MAPS-based tracking detector, consisting of 7 concentric layers, during its Inner Tracking System (ITS) upgrade during the second long shutdown of the LHC, denoted ITS2 \cite{ALICE:2013nwm}, in the first adoption of MAPS at the LHC. 
To further improve tracking and vertexing performance, the ITS3 upgrade \cite{The:2890181} seeks to replace the three innermost layers of the ITS2 with three fully cylindrical layers consisting of wafer-scale bent sensors, targeting an excellent spatial resolution of the chips (\SI{5}{\micro\meter}) and  low material budget (0.09\% X$_{0}$/layer). 
The layers will rely on the stitching technique supported in the commercially available TPSco 65 nm CMOS imaging process. 
In the context of the ITS3 upgrade, a set of small-scale test structures were produced in a submission in December 2020: Analogue Pixel Test Structure (APTS \cite{Rinella:2024jzi}), Digital Pixel Test Structure (DPTS \cite{Rinella:2022htp}), and the Circuit Exploratoire 65 nm (CE-65 \cite{Bugiel:2022gwe}). 
A second submission at the beginning of 2023 allowed for the evolution of some of the small-scale test structures, including the CE-65v2: the evolution of CE-65.

The CE-65v2 chip was developed to investigate the charge collection and electrical properties of the 65 nm CMOS process through a selection of chip variants.
The chip consists of a matrix of 1152 pixels organised in 48 columns and 24 rows, with a rolling-shutter readout. 
The in-pixel electronics consists of a AC-coupled amplifier that is DC-separated from the input stage of the readout electronics, allowing for the application of a resetting voltage for the reverse biasing of the sensor to the reset node. A resetting voltage of 10 V was applied throughout these studies in order to achieve full depletion.
In total, there are 15 variants of the CE-65v2 chip targeting the exploration of 3 main axes:
\begin{itemize}
    \item Process variation: Standard, Modified, Modified with Gap
    \item Pitch variation: \SI{15}{\micro\meter}, \SI{18}{\micro\meter}, \SI{22.5}{\micro\meter}
    \item Matrix geometry: square vs hexagonal
\end{itemize}
In this work, the first two axes are explored. A total of four chips corresponding to the \SI{15}{\micro\meter} and \SI{22.5}{\micro\meter} chips in the Standard process and the \SI{15}{\micro\meter} and \SI{22.5}{\micro\meter} chips in the Modified with Gap process were considered in these studies. 

Adopting a smaller pitch allows for an improvement in spatial resolution, but comes with significant drawbacks with respect to power consumption, readout-rate, as well as difficulties during manufacturing. 
Thus it becomes necessary to explore the pitch size at which the desired spatial resolution can be achieved, without compromising other aspects of the chip and system design.

\begin{figure}[htbp]
\centering
\includegraphics[width=.48\textwidth]{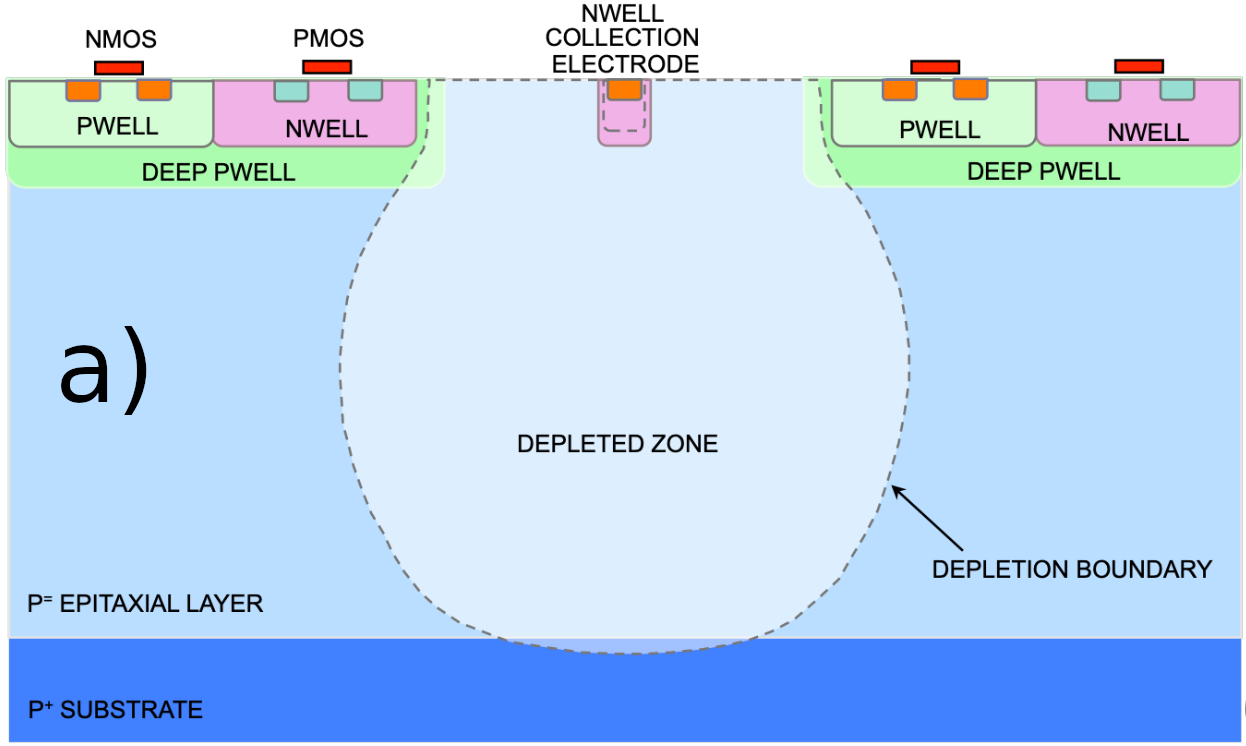}
\hfill
\includegraphics[width=.48\textwidth]{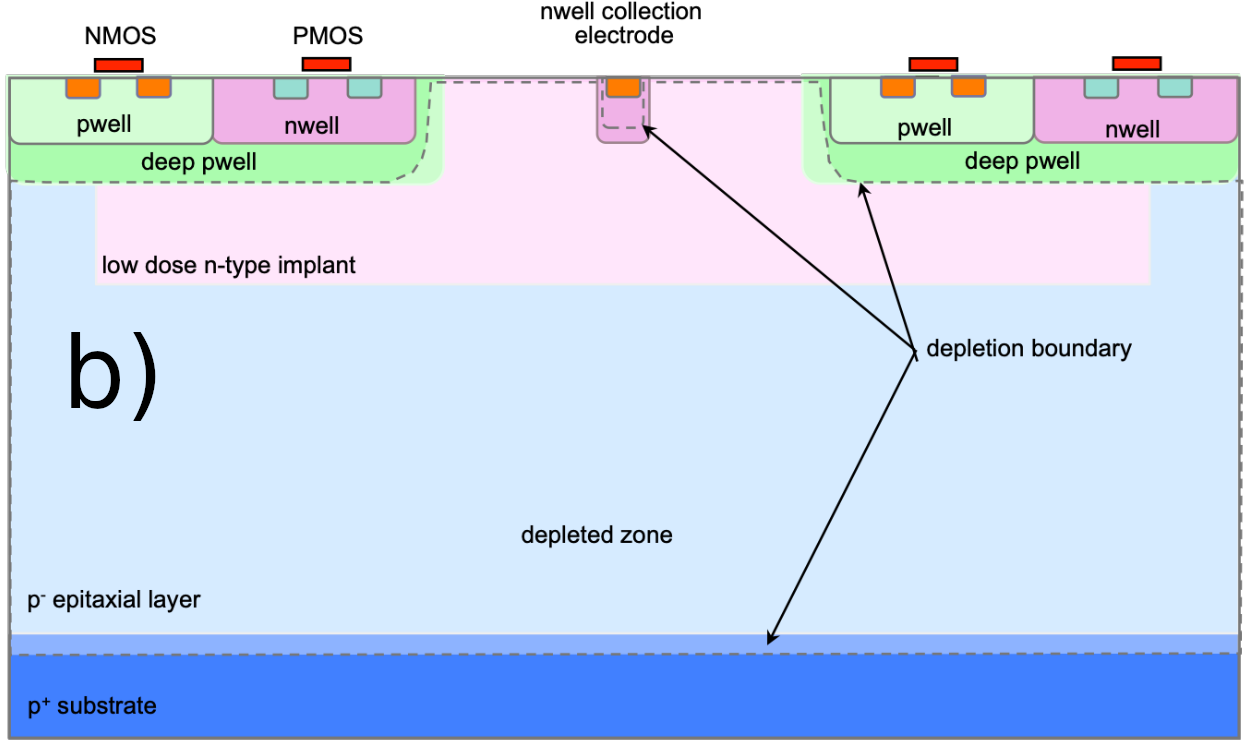}
\caption{Cross-sections of the CE-65v2 pixels detailing the Standard process (a) and the Modified with Gap process (b). The figures were adapted from Ref. \cite{The:2890181}. \label{fig:cross-section}}
\end{figure}

The process variations are detailed in Ref. \cite{Snoeys:2280552}. 
Figure \ref{fig:cross-section}a depicts the Standard process consisting of an n-well collection electrode, and in-pixel CMOS circuitry that is isolated from the epitaxial layer by a deep p-well. 
The depletion region begins to develop at the n-well collection electrode and follows a balloon shape in the epitaxial layer until the p+ substrate is reached. 
Due to the limited depth of the epitaxial region of $\mathcal{O}$(\SI{10}{\micro\meter}), the lateral region remains undepleted, resulting in diffusion-dominated charge collection. 
In effect, charge collection is slow and subject to charge trapping, whilst exhibiting high charge sharing between pixels. 
Figure \ref{fig:cross-section}b depicts the Modified with Gap process which has, in addition, a deep low-dose n-type implant between the epitaxial layer and the CMOS circuitry, with gaps at the pixel edges. 
The depletion region in the Modified with Gap process extends laterally, as the gaps allow for the development of the electric field. 
The lateral electric field induces drift-dominated charge collection even at the edges, resulting in faster charge collection and reduced charge sharing.

\section{Characterisation with radioactive source}
\label{sec:lab}
 
\begin{figure}[htbp]
\centering
\includegraphics[width=.48\textwidth]{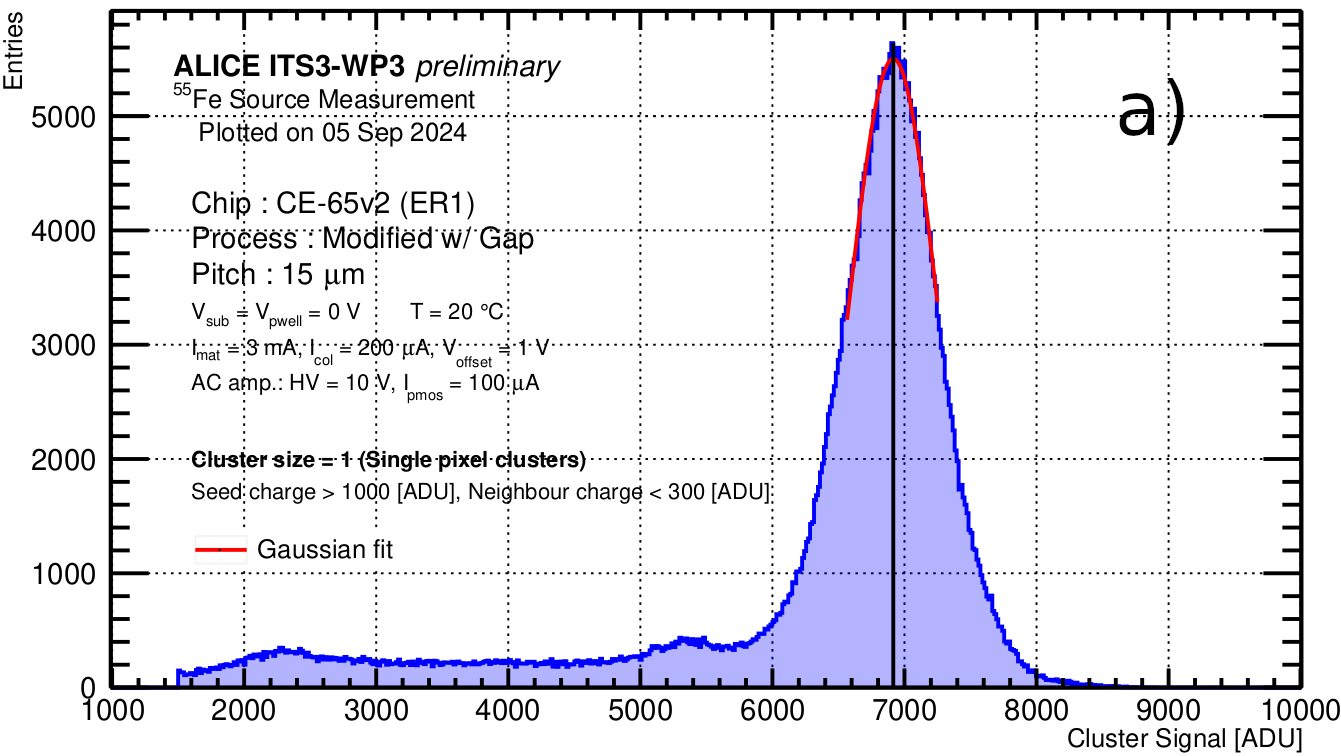}
\hfill
\includegraphics[width=.48\textwidth]{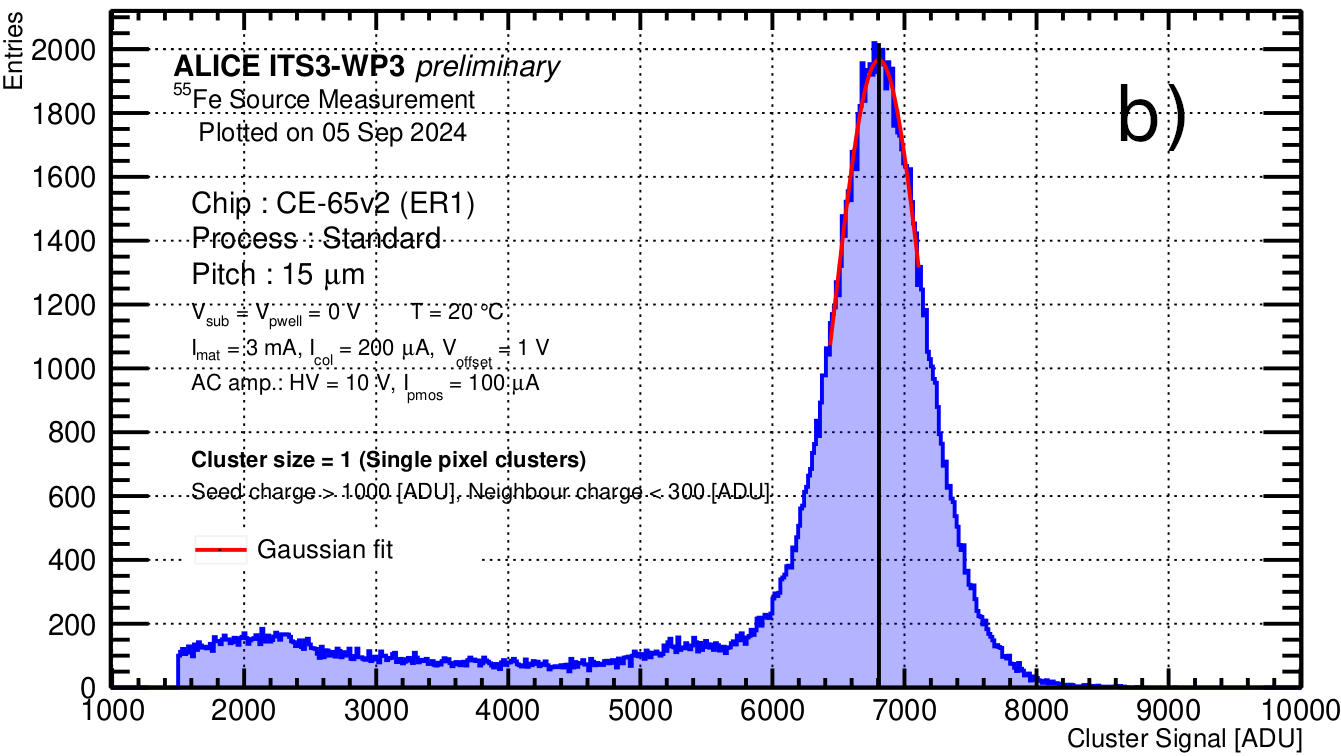}
\caption{$^{55}$Fe spectrum for single pixel clusters of the entire matrix of the CE-65v2 chip in the Modified with Gap process (a) and Standard process (b) with a \SI{15}{\micro\meter} pitch. The Gaussian fit is marked in red.\label{fig:fe55_spectrum}}
\end{figure}

\begin{figure}[htbp]
\centering
\hspace{-2.7em}
\includegraphics[width=.48\textwidth]{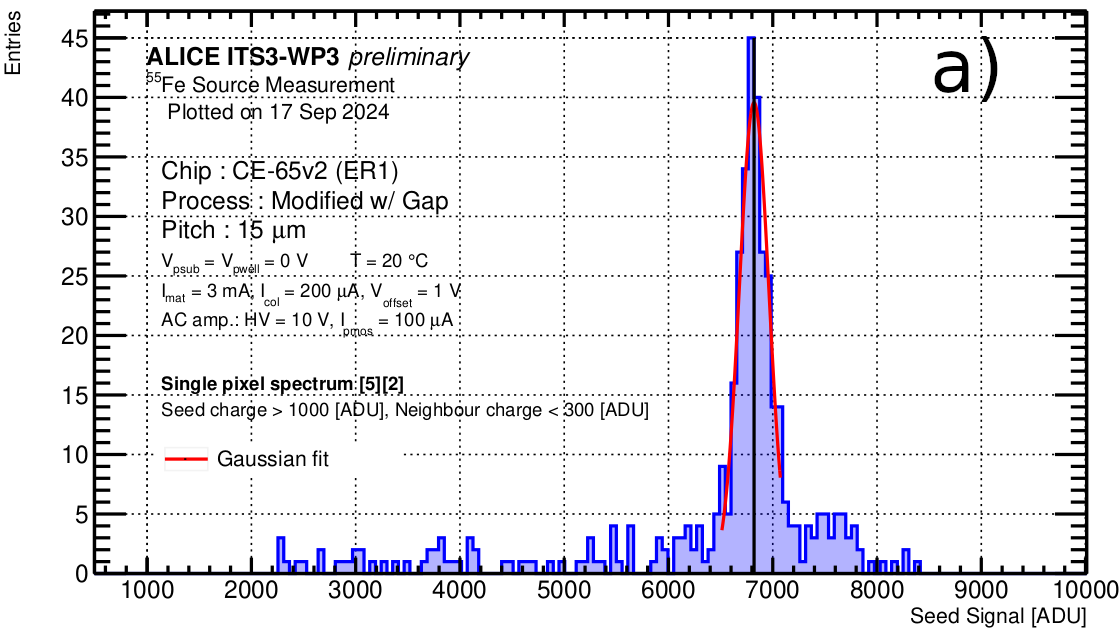}
\hspace{-0.5em}
\includegraphics[width=.58\textwidth]{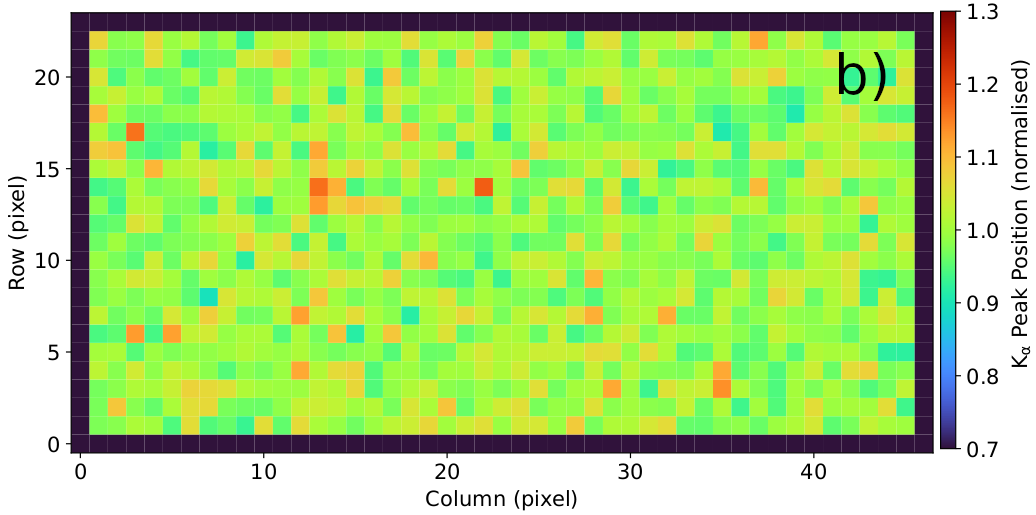}
\caption{a) $^{55}$Fe spectrum for single pixel clusters for an individual pixel (x=5, y=2) of the CE-65v2 chip in the Modified with Gap process with a \SI{15}{\micro\meter} pitch. The Gaussian fit is marked in red. b) Gain distribution of the CE-65v2 chip in the Modified with Gap process with a \SI{15}{\micro\meter} pitch obtained by considering the normalized K$_{\alpha}$ peak position (Figure \ref{fig:gain_2d}a) for each pixel.\label{fig:gain_2d}}
\end{figure}

An extensive characterisation of the CE-65v2 chip was performed using X-rays from an $^{55}$Fe source. 
In particular, the X-ray spectrum of $^{55}$Fe is determined by the electron transitions possible during the electron capture decay of $^{55}$Fe to $^{55}$Mn. 
The most prominent peak (K$_{\alpha}$) is at a known energy of 5.9 keV, with a secondary peak (K$_{\beta}$) at an energy of 6.5 keV, with the peak width being determined primarily by the energy resolution of the given detector. 
A mapping from the measured Analog-to-Digital Units (ADUs), resulting from the quantization of the measured voltage by a 16-bit Analog-to-Digital Converter, to energy can be performed by matching the peak position of the K$_{\alpha}$ (and K$_{\beta}$) X-rays.
The energy is often expressed in terms of electrons corresponding to the number of electron-hole pairs produced by the impinging X-ray. 
The different flavours of the CE-65v2 were measured at constant 20$^{o}$C using a chiller for temperature control. 
The signal is computed by subtracting temporally consecutive pixel frames to minimize baseline noise fluctuations.
For a given event the $3\times3$ matrix surrounding the most energetic pixel is considered. 
If the pixel passes a threshold of 1000 ADUs then it is considered a seed for cluster reconstruction. 
Adjacent pixels must pass a threshold of 300 ADUs, corresponding to approximately 2 times the Root Mean Square (RMS) noise, to be considered neighbours. 
Figure \ref{fig:fe55_spectrum}a depicts the measured $^{55}$Fe spectrum for single pixel clusters for the CE-65v2 chip in the Modified
with Gap process with a \SI{15}{\micro\meter} pitch. 
The main K$_{\alpha}$ peak, centered around $\sim$6900 ADUs, is fitted with a Gaussian distribution to extract the peak position. 
The neighbouring K$_{\beta}$ peak, which would be expected at $\sim$7480 ADUs, cannot be resolved due to pixel-to-pixel gain variations across the matrix. 
The same procedure was repeated for the \SI{22.5}{\micro\meter} pitch chip, as well as the \SI{15}{\micro\meter} and \SI{22.5}{\micro\meter} chips in the Standard process, yielding similar spectra and peak positions. As shown in Figure \ref{fig:fe55_spectrum}b, the main K$_{\alpha}$ peak position for the \SI{15}{\micro\meter} Standard process chip is within 1.6\% of the \SI{15}{\micro\meter} Modified with Gap process chip. This outcome is anticipated, as the capacitance should not vary significantly when the chips are fully depleted.
The same procedure was repeated for the single-pixel spectrum of individual pixels, yielding much cleaner spectra, at the cost of limited statistics, as can be seen in Figure \ref{fig:gain_2d}a. 

The energy resolution of 4.8\%, obtained as the ratio of the Full Width
at Half Maximum and mean peak position, is considerably better than that of the global spectrum in Figure \ref{fig:fe55_spectrum}a, where it is 11.5\%. 
Moreover, the K$_{\beta}$ peak can be clearly resolved. %, and is fit with a \textbf{double-Gaussian}. 
Figure \ref{fig:gain_2d}b summarizes the gain uniformity of the \SI{15}{\micro\meter} chip in the Modified with Gap process by depicting the K$_{\alpha}$ peak position for each individual pixel, normalized by the mean K$_{\alpha}$ peak position for all pixels. 
No clear spatial pattern can be discerned, and indeed the variance of the gain between pixels is of $\mathcal{O}$(5\%). 
Similar trends hold for the other three chips, with the variance being of $\mathcal{O}$(5\%) in all cases. 
The initial characterisation with radioactive sources provided valuable insights into the gain uniformity and baseline performance, which were essential for validating the test beam measurements.

%\begin{figure}[htbp]
%\centering
%\includegraphics[width=.58\textwidth]{Figures/gain_2d_gap15sq.png}
%\caption{Gain distribution of the CE-65v2 chip in the Modified with Gap process with a \SI{15}{\micro\meter} pitch obtained by considering the normalized K$_{\alpha}$ peak position in $^{55}$Fe source measurements depicted in Figure \ref{fig:fe55_spectrum}b.\label{fig:gain_2d}}
%\end{figure}

\section{Testbeam results}
\label{sec:testbeam}

Test beam measurements were conducted at CERN SPS (H6 beamline) \cite{Banerjee:2774716} to study the efficiency and resolution of different CE-65v2 variants. 
A telescope consisting of six ALPIDE \cite{Mager:2016yvj} planes and using the DPTS \cite{Rinella:2022htp} chip as a trigger was used. 
The telescope resolution was estimated to be \SI{2.2}{\micro\meter} using a telescope optimizer \cite{telescopeoptimizer} for the given geometry. 
All tests were conducted at 20$^{o}$C using a chiller to stabilize the device under testing.  
Data analysis was carried out using the Corryvreckan framework \cite{Dannheim:2020jlk}, with tracks reconstructed using a straight-line track model that required hits on all six ALPIDE planes. 
Clusters on the CE-65v2 chip were built by summing all pixels in a $3\times3$ window around a seed pixel passing an 100 e$^{-}$ threshold. 
The maximum cluster size is thus by definition 9.
For electron threshold scans, identical seed and neighbour threshold cuts were applied for the pixels in the cluster, before their contributions were summed.
A signal-to-noise ratio greater than 3 was required for all seed pixels.
Multiple cluster candidates were considered corresponding to each pixel passing the required seed e$^{-}$ threshold and signal-to-noise ratio.
Clusters within a \SI{75}{\micro\meter} radius were associated with a given track, with the nearest cluster selected in cases of multiple candidates.

\begin{figure}[htbp]
\centering
\includegraphics[width=.48\textwidth]{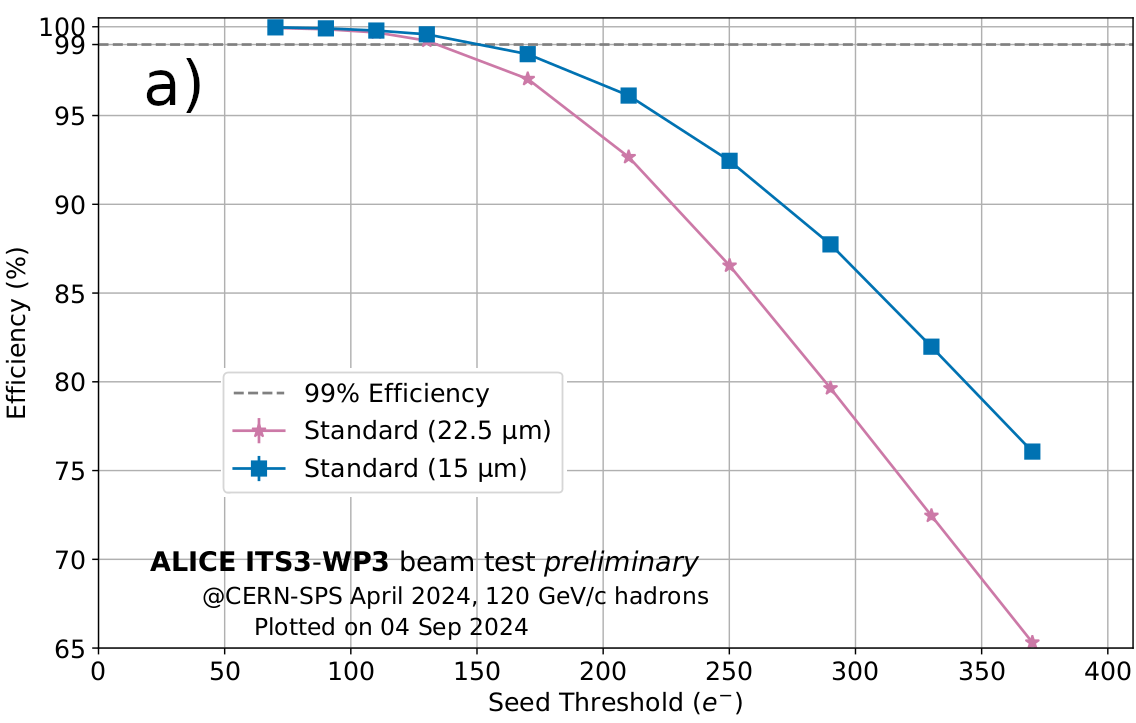}
\hfill
\includegraphics[width=.48\textwidth]{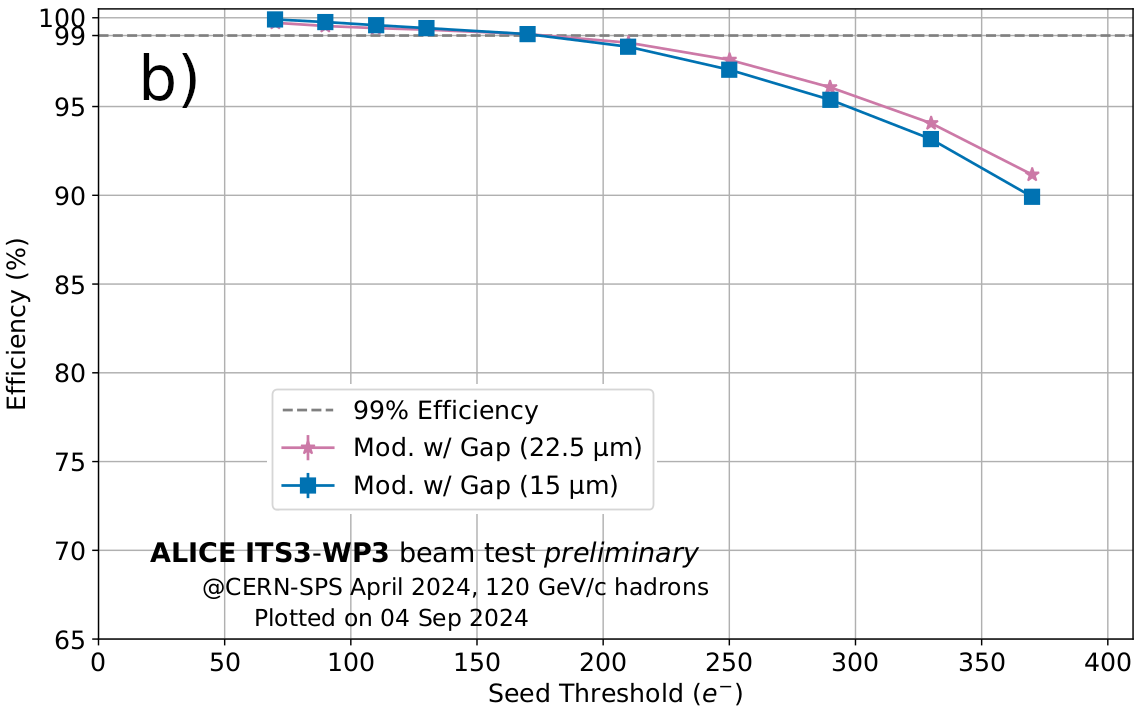}
\caption{Efficiency as a function of seed threshold (e$^{-}$) for the CE-65v2 chip with \SI{15}{\micro\meter} (blue) and a \SI{22.5}{\micro\meter} pitch (pink) in the Standard (a) and the Modified with Gap process (b). The 99\% efficiency line (dotted) is marked explicitly.\label{fig:efficiency}}
\end{figure}

The efficiency as a function of electron threshold was studied for each of the four chip variants.
In the Standard process, an efficiency of over 99\% is achieved up to $\sim$130 e$^{-}$ and $\sim$150 e$^{-}$ for the \SI{15}{\micro\meter} and \SI{22.5}{\micro\meter} chips respectively, as depicted in Figure \ref{fig:efficiency}a. 
The efficiency drops with respect to increasing threshold, but does so faster for the \SI{22.5}{\micro\meter} pitch chip.
In the Modified with Gap process, an efficiency of over 99\% is achieved up to $\sim$180 e$^{-}$ for both the \SI{15}{\micro\meter} and \SI{22.5}{\micro\meter} chips, as depicted in Figure \ref{fig:efficiency}b. 
The drop in efficiency is considerably slower with respect to the Standard process chips. 
The difference with respect to pitch size is marginal, with the \SI{22.5}{\micro\meter} chip maintaining slightly higher efficiency.

\begin{figure}[htbp]
\centering
\includegraphics[width=.48\textwidth]{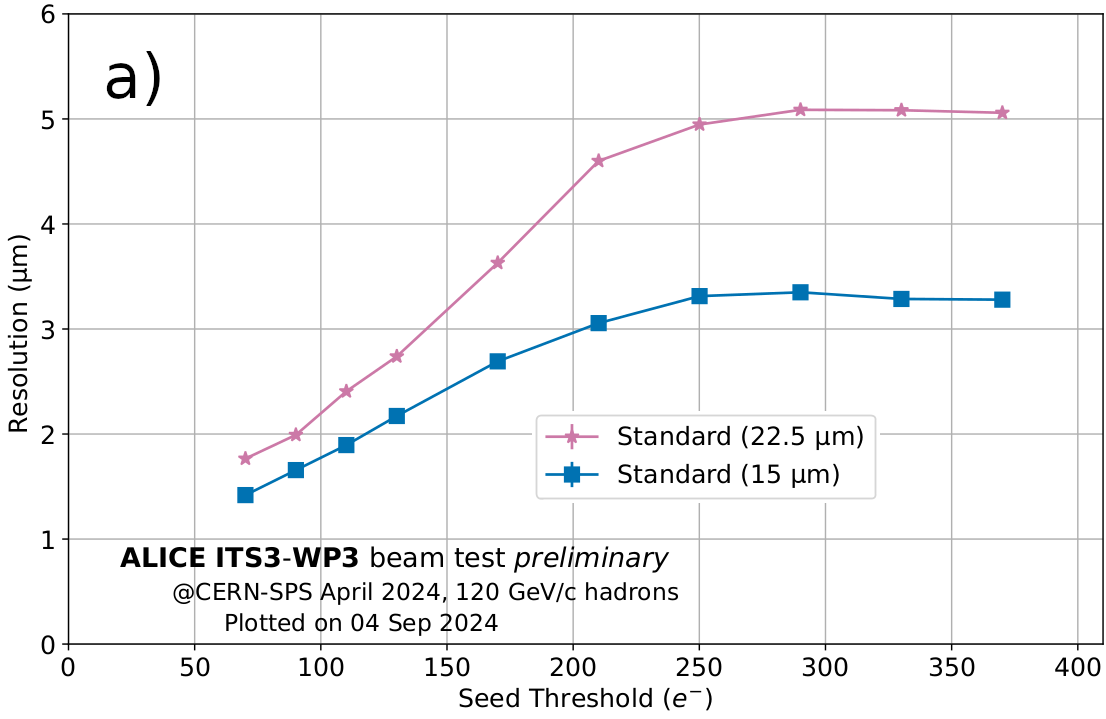}
\hfill
\includegraphics[width=.48\textwidth]{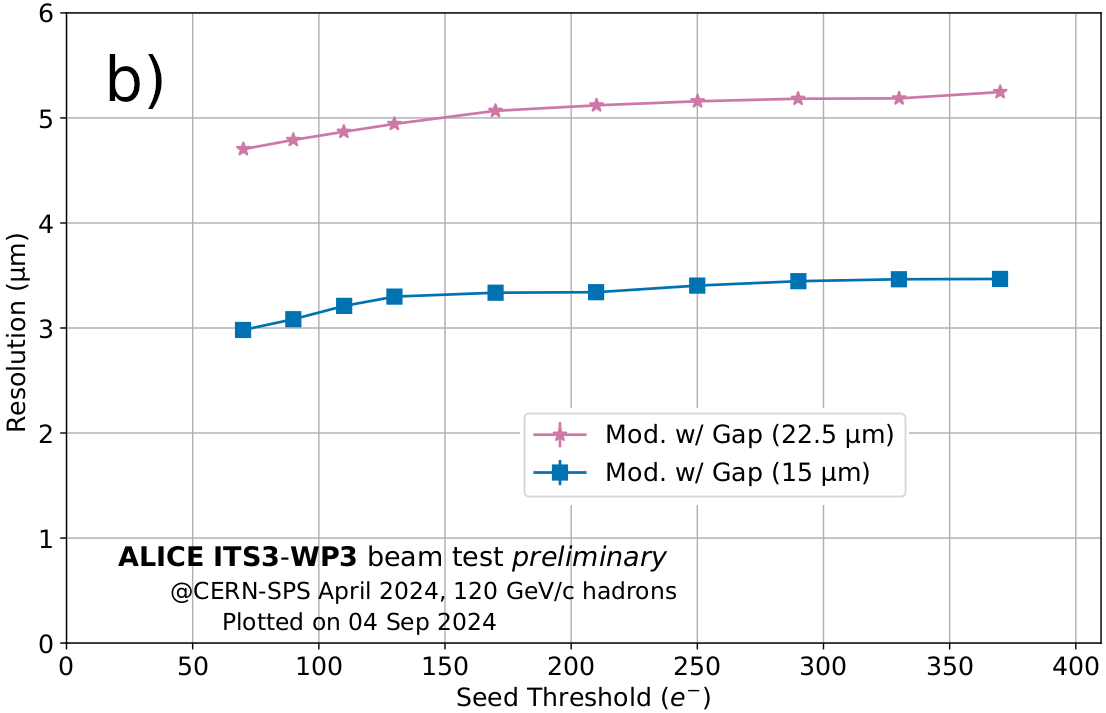}
\caption{Resolution as a function of seed threshold (e$^{-}$) for the CE-65v2 chip with \SI{15}{\micro\meter} (blue) and a \SI{22.5}{\micro\meter} pitch (pink) in the Standard (a) and the Modified with Gap process (b).\label{fig:resolution}}
\end{figure}

The spatial resolution of the chip variants was likewise studied with respect to increasing electron threshold. 
For each threshold the full analogue information was used to compute the  centre-of-mass position of a given cluster using the seed and neighbour thresholds given previously in this Section.
In the Standard process, an excellent resolution of $\sim$\SI{1.5}{\micro\meter} and $\sim$\SI{2}{\micro\meter} can be achieved at a seed threshold of 70 e$^{-}$ for the \SI{15}{\micro\meter} and \SI{22.5}{\micro\meter} chips, as depicted in Figure \ref{fig:resolution}a.
The performance degrades quickly in the 70 e$^{-}$ - 250 e$^{-}$ range as the cluster size decreases, before plateauing at $\sim$\SI{3}{\micro\meter} and $\sim$\SI{4}{\micro\meter} for the \SI{15}{\micro\meter} and \SI{22.5}{\micro\meter} chips.
For both chips a resolution that is considerably better than the binary resolution of pitch/$\sqrt{12}$ is achieved.
Figure \ref{fig:resolution}b depicts the resolution as a function of increasing e$^{-}$ threshold for the Modified with Gap process chips. The Modified with Gap chips achieve a worse resolution compared to the Standard process chips for all considered thresholds, especially for the \SI{15}{\micro\meter} pitch chip. The resolution is, however, more stable with increasing thresholds. Comparing the two pitches for the Modified with Gap process chips, the \SI{22.5}{\micro\meter} chip displays less variation increasing slightly to just above \SI{5}{\micro\meter} resolution, in parity with the Standard process chip.

\begin{figure}[htbp]
\centering
\includegraphics[width=.48\textwidth]{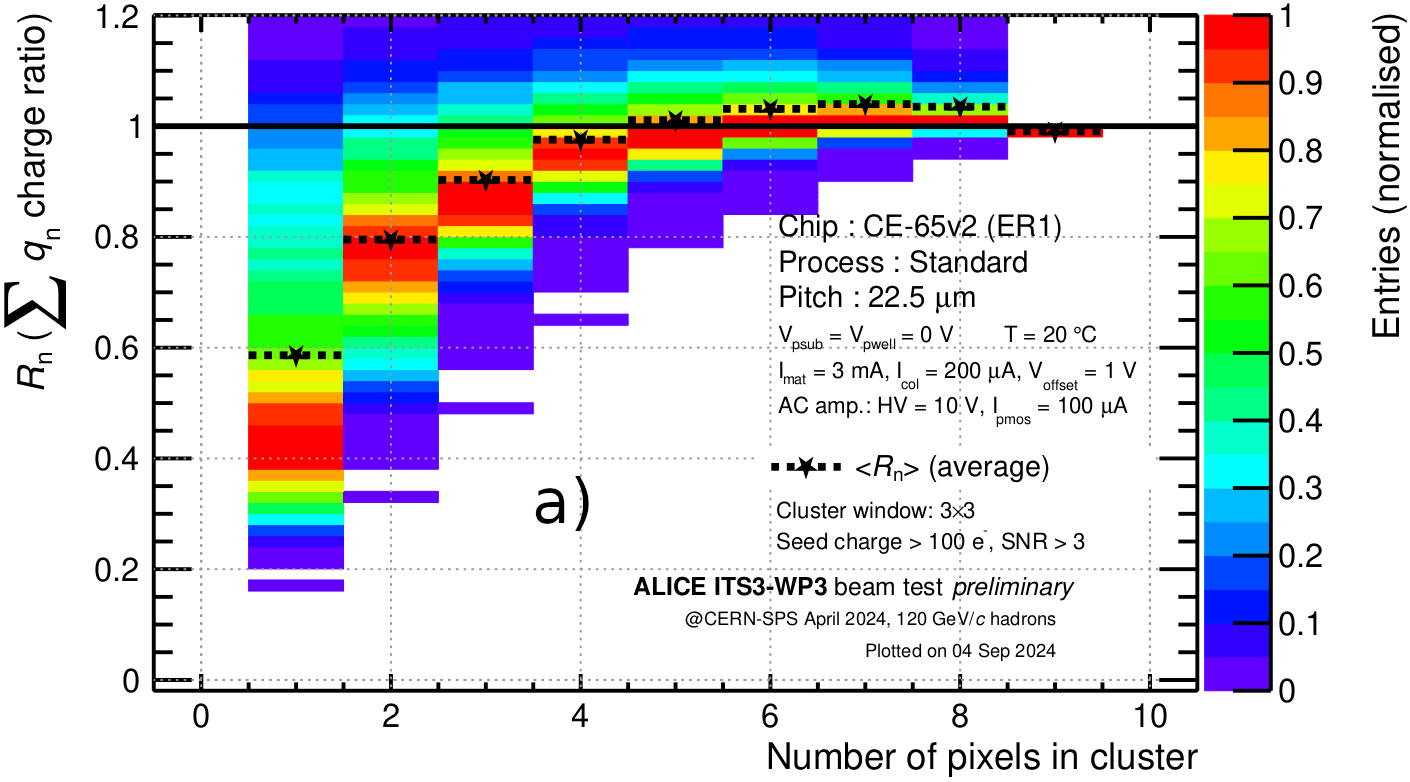}
\hfill
\includegraphics[width=.48\textwidth]{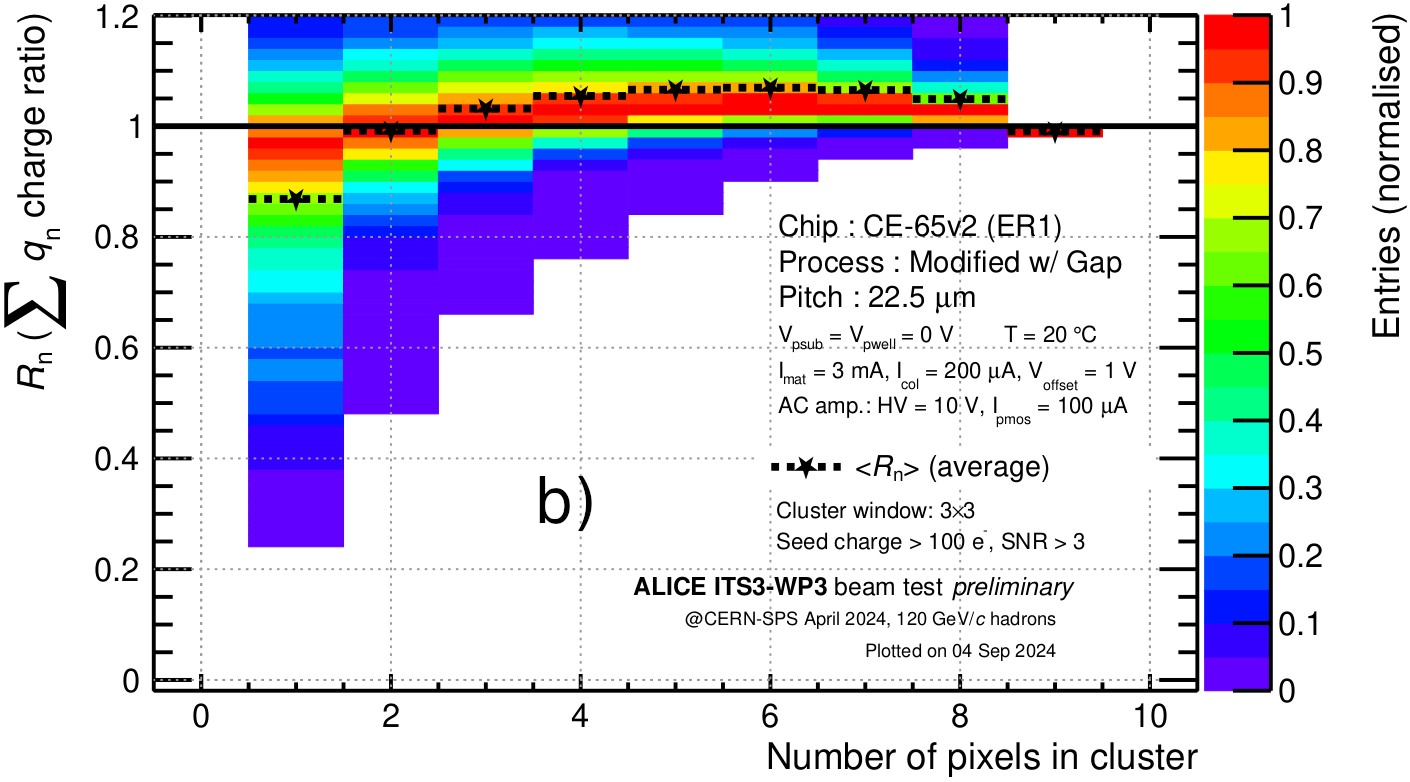}
\caption{a) Accumulated charge ratio as a function of number of pixels in a cluster for the CE-65v2 chip with a \SI{22.5}{\micro\meter} in the Standard (a) and Modified with Gap process (b). \label{fig:charge_shape}}
\end{figure}

Charge sharing behavior significantly impacts both efficiency and spatial resolution, making it essential to assess across different pixel configurations.
To study the extent of charge sharing for the different CE-65v2 variants, the accumulated charge ratio was determined by charge-ordering the pixels in a given cluster, and successively summing the charge of individual pixels, until a cluster size of 9 was obtained. 
The accumulated charge was then normalized by the total charge, defined as the sum of the charge of all pixels in a 3x3 window around the seed pixel, to obtain the accumulated charge ratio.
Figure \ref{fig:charge_shape}a depicts the accumulated charge ratio as a function of pixels in the cluster for the \SI{22.5}{\micro\meter} pitch chip in the Standard process. 
The average charge fraction carried by the central pixel is less than $\sim$60\%, with the most probable value being even lower at $\sim$45\%, showing the high charge sharing characteristic of the Standard process. 
The most extreme charge sharing was found for the \SI{22.5}{\micro\meter} pitch chip in the Standard process due to the competing effects of a larger pitch, whereby more charge is collected due to the larger area, and the electric field not propagating well at the pixel edges, resulting in considerably less charge drift.
It is important to note that due to charge ordering, negative noise contributions, resulting from the frame subtraction described in Section \ref{sec:lab}, can appear in the larger cluster size bins, as shown on the right side of Figure \ref{fig:charge_shape}a. These noise contributions may cause the accumulated charge ratio to exceed 100\%.

Figure \ref{fig:charge_shape}b, on the other hand, depicts the accumulated charge ratio as a function of pixels in the cluster for the \SI{22.5}{\micro\meter} pitch chip in the Modified with Gap process. 
The average charge fraction carried by the central pixel is considerably higher with respect to the Standard process chip, with the charge fraction being over 85\%. 
When considering the two most energetic pixels, both the average and most probable value of accumulated charge are around unity, suggesting that the vast majority of events consist of single-pixel or two-pixel events. 
Indeed, it is to be expected that if a hit occurs close to the edge of the pixel, then charge is shared with the neighbouring pixel, otherwise it is almost entirely collected due to the strong drift current evident in the Modified with Gap process.
Due to the Gap modification at the pixel boundaries, the electric field does propagate well at the pixel edges, unlike in the Standard process. 
Coupled with a larger pitch this results in the lowest charge sharing of the four chip variants that were studied. 

\section{Conclusion}
\label{sec:conclusion}

This work presents the first results of the CE-65v2 chip.
A detailed comparison of the Standard process and the Modified with Gap process, as well as varying pixel pitch (\SI{15}{\micro\meter}, \SI{22.5}{\micro\meter}) was made with particular emphasis on charge sharing properties and their ramifications on efficiency and resolution. 
Gain uniformity up to the $\mathcal{O}$(5\%) level was demonstrated for the considered CE-65v2 variants in $^{55}$Fe lab tests.
An excellent resolution in large-matrix 65 nm CMOS test structures was obtained during beam test measurements at the CERN SPS. The sub \SI{3}{\micro\meter} spatial resolution obtained in the Standard process for both pitches satisfies FCC-ee requirements \cite{Barchetta:2021ibt}, and allows tradeoffs in pixel pitch with respect to power consumption, readout-rate, and manufacturing-ease. This enables a wide range of applications for 65 nm TPSCo MAPS.
However, the spatial resolution was observed to quickly degrade when the chip is operated outside the low electron threshold regime, making it particularly sensitive to noise and radiation defects, including charge trapping.
The Modified with Gap process displays a wide operating range with over 99\% efficiency being achievable up to $\sim$180 e$^{-}$ for both the \SI{15}{\micro\meter} and \SI{22.5}{\micro\meter} chip. The wider range of operation and faster charge collection makes the process more suited for high radiation environments.

The complete characterisation of the CE-65v2 chip is currently underway.
A full analysis of the CERN SPS results including the Modified “intermediate” process modification and \SI{18}{\micro\meter} pitch chips is in progress. 
In addition, the use of the resetting voltage as a tool to tune charge sharing, and the impact of a hexagonal matrix pixel arrangement on cluster size and resolution are likewise being studied. 
Radiation tolerance studies using irradiated CE-65v2 chips were performed in May 2024 at DESY, and will complement the results obtained thus far. In summary, the characterisation of the CE-65v2 chip has supplemented the APTS \cite{Rinella:2024jzi} and DPTS \cite{Rinella:2022htp} studies in the validation of the 65 nm TPSCo process as a candidate technology for advanced particle detection applications, including the ALICE ITS3 upgrade. The demonstrated spatial resolution, efficiency, gain uniformity, and ongoing studies pertaining to matrix geometry and radiation tolerance will further help refine the design parameters necessary for efficient tracking systems in future high-energy physics experiments.

\acknowledgments
The measurements leading to these results were performed at the Test
Beam Facilities at CERN (Switzerland). This project has received funding from the European Union’s Horizon 2020 Research and Innovation Programme under Grant Agreements 101004761 (AIDAinnova) and 101057511 (EURO-LABS).

%\paragraph{Note added.} This is also a good position for notes added
%after the paper has been written.

% Bibliography

%% [A] Recommended: using JHEP.bst file
\bibliographystyle{JHEP}
\bibliography{biblio.bib}

\end{document}